\documentclass[aps,prl,reprint,superscriptaddress,showpacs,floatfix,longbibliography]{revtex4-1}

\usepackage{graphicx}
\usepackage{amssymb,amsmath}
\usepackage{dcolumn}
\usepackage{bm}
\usepackage[mathlines]{lineno}
\usepackage{soul,color}
\usepackage{multirow}
\usepackage{amssymb,amsmath,amsthm}

\begin{document}

\title{Long-wavelength deformations and vibrational modes
       in empty and\\
       liquid-filled microtubules and nanotubes:
       A theoretical study}

\author{Dan~Liu}
\affiliation{Physics and Astronomy Department,
             Michigan State University,
             East Lansing, Michigan 48824, USA}

\author{Arthur~G.~Every}
\affiliation{School of Physics,
             University of the Witwatersrand,
             Private Bag 3,
             2050 Johannesburg,
             South Africa
}

\author{David Tom\'{a}nek}
\email
            {tomanek@pa.msu.edu}%
\affiliation{Physics and Astronomy Department,
             Michigan State University,
             East Lansing, Michigan 48824, USA}

\date{\today} 

\begin{abstract}
We propose a continuum model to predict long-wavelength
vibrational modes of empty and liquid-filled tubules that are very
hard to reproduce using the conventional force-constant matrix
approach based on atomistic {\em ab initio} calculation.
We derive simple quantitative expressions for long-wavelength
longitudinal and torsional acoustic modes, flexural acoustic
modes, as well as the radial breathing mode of empty or
liquid-filled tubular structures that are based on continuum
elasticity theory expressions for a thin elastic plate. We
furthermore show that longitudinal and flexural acoustic modes of
tubules are well described by those of an elastic beam resembling
a nanowire. Our numerical results for biological microtubules and
carbon nanotubes agree with available experimental data.
\end{abstract}



\maketitle

\section{Introduction}

Tubular structures with diameters ranging from nanometers to
meters abound in nature
to fill various functions. The elastic response of most tubular
structures is dominated by low-frequency flexural acoustic (ZA)
modes. Much attention has been devoted to the nanometer-wide
carbon nanotubes (CNTs)~\cite{DT188}, which are extremely
stiff~\cite{DT059}, and to their flexural
modes~\cite{{DT059},{Sazonova},{vibrationC},{Gibson20071}}.
Correct description of soft flexural modes in stiff quasi-1D
systems like nanotubes and nanowires is essential for calibrating
nanoelectromechanical systems used for ultrasensitive mass
detection and radio-frequency signal
processing~\cite{{Roukes},{Sazonova}}. In CNTs and in related
graphene nanoribbons, flexural ZA modes have also been shown to
significantly influence the unsurpassed lattice thermal
conductivity~\cite{flex-graphene16}. Much softer microtubules
formed of tubulin proteins, with a diameter $d{\approx}20$~nm, are
key components of the cytoskeleton and help to maintain the shape
of cells in organisms. In spite of their importance, there are
only scarce experimental data available describing the elastic
behavior of microtubules.
The conventional approach to calculate the frequency spectrum is
based on an atomistic calculation of the force-constant matrix.
This approach often fails for long-wavelength acoustic modes, in
particular the soft flexural ZA modes, due to an excessive demand
on supercell size and basis convergence. Typical results of this
shortcoming are numerical artifacts such as imaginary vibration
frequencies~\cite{Ling10}.


Here we offer an alternative way, based on continuum elasticity
theory~\cite{Love-book} and its extension to
planar~\cite{DT255} and tubular structures~\cite{{QianJAP07},%
{SirenkoPRE96},{WangMicrotubules}}, to predict the frequency of
acoustic modes in quasi-1D structures such as empty and
liquid-filled tubes consisting of stiff graphitic carbon or soft
tubulin proteins.
While the scope of our approach is limited to long-wavelength
acoustic modes, the accuracy of vibration frequencies calculated
using the simple expressions we derive surpasses that of
conventional atomistic {\em ab initio} calculations. Our approach
covers longitudinal and torsional modes, flexural modes, as well
as the radial breathing mode. We show that longitudinal and
flexural acoustic modes of tubules are simply related to those of
an elastic beam resembling a nanowire. Since the native
environment of tubulin nanotubes contains water, we specifically
consider the effect of a liquid on the vibrational modes of
tubular structures. Our numerical results for tubulin microtubules
and carbon nanotubes agree with available experimental data.

%

\begin{figure}[b]
\includegraphics[width=1.0\columnwidth]{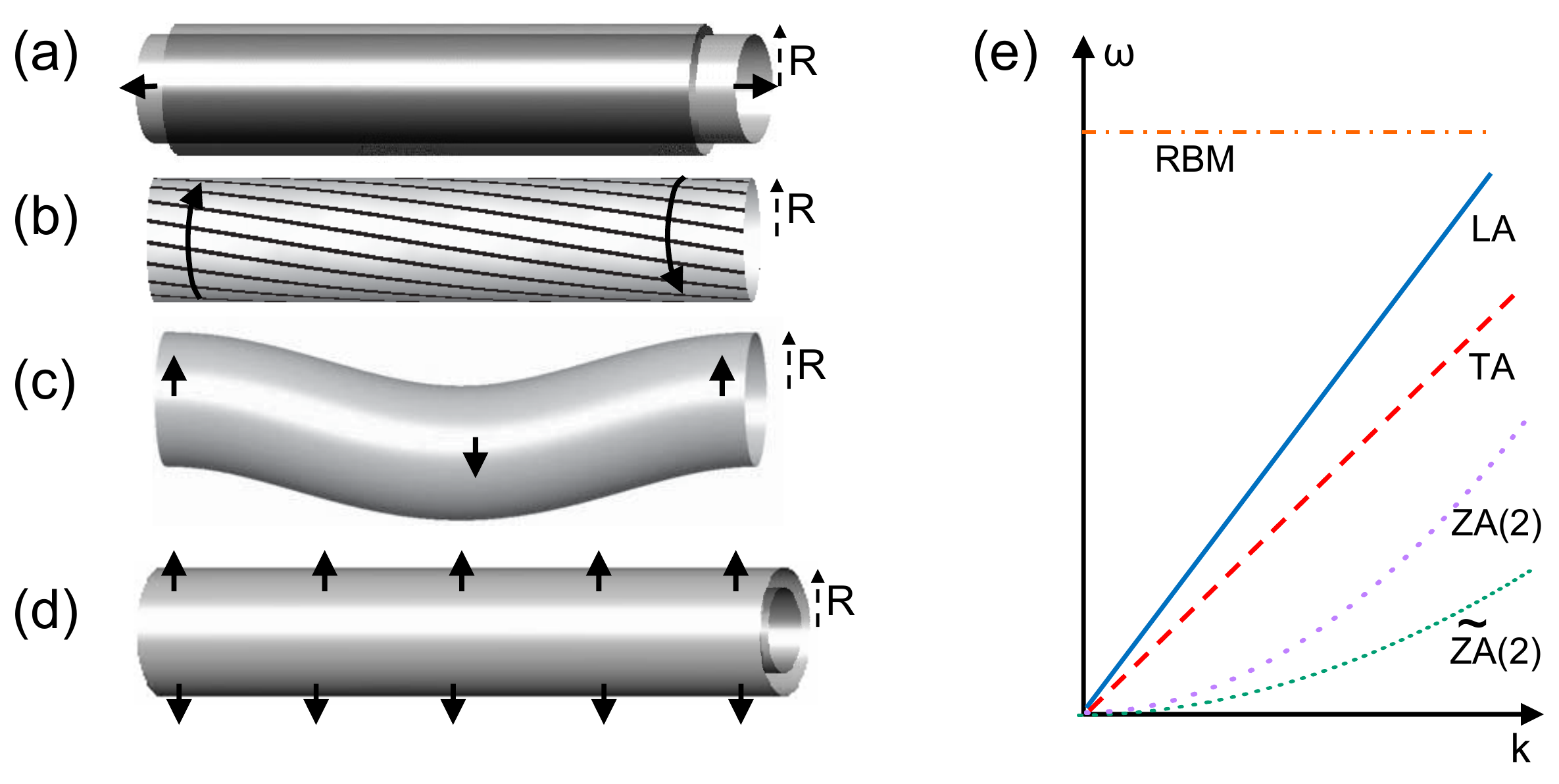}
\caption{(Color online) Schematic representation of important
deformations of a tubular structure. (a) Longitudinal acoustic
(LA, stretching), (b) torsional acoustic (TA, torsion), (c)
flexural acoustic (doubly degenerate ZA, bending), and (d) the
radial breathing mode (RBM). (e) Schematic dispersion relations of
the corresponding long-wavelength phonon modes. The tilde denotes
liquid-filled tubules.%
\label{fig1} }
\end{figure}


\section{Continuum elasticity approach}

A 1D tubular structure of radius $R$ can be thought of as a
rectangular 2D plate of width 2$\pi$R that is rolled up to a
cylinder. Consequently, the elastic response of 1D tubules to
strain, illustrated in Figs.~\ref{fig1}(a)-\ref{fig1}(d), is
related to that of the constituting 2D plate. To describe this
relationship in the linear regime and calculate the frequency of
long-wavelength vibrational modes in 1D tubular structures, we
take advantage of a continuum elasticity formalism that has been
successfully adapted to 2D structures~\cite{DT255}.

As shown earlier~\cite{DT255}, elastic in-plane deformations of a
plate of indefinite thickness may be described by the
$(3{\times}3)$ 2D elastic stiffness matrix, which is given in
Voigt notation by
\begin{eqnarray}
\left( \begin{array}{ccc}
 c_{11} & c_{12} & 0 \\
 c_{12} & c_{22} & 0 \\
  0     &   0    & c_{66}
       \end{array}
\right) \,.%
\label{Eq1}
\end{eqnarray}
Resistance of such a plate to bending is described by the flexural
rigidity $D$. For a plate suspended in the $x-y$ plane, $c_{11}$
and $c_{22}$ describe the longitudinal strain-stress relationship
along the $x-$ and $y-$direction, respectively. $c_{66}$ describes
the elastic response to in-plane shear. For an isotropic plate,
which we consider here, $c_{11}=c_{22}$,
$c_{66}=(c_{11}-c_{12})/2$, and the in-plane Poisson ratio
${\alpha}=c_{12}/c_{11}$. Considering a 3D plate of finite
thickness $h$, characterized by the $(6{\times}6)$ elastic
stiffness matrix $C_{ij}$, the coefficients of the 2D elastic
stiffness matrix $c_{ij}$ for the equivalent plate of indefinite
thickness are related by
$c_{ij}=h{\cdot}C_{ij}(1-\alpha_{\perp}^2)$. This expression takes
account of the fact that stretching a finite-thickness slab of
isotropic material not only reduces its width by the in-plane
Poisson ratio $\alpha$, but also its thickness by the out-of-plane
Poisson ratio $\alpha_{\perp}$. This consideration is not needed
for layered compounds such as graphite, where the inter-layer
coupling is weak and $\alpha_{\perp}{\approx}0$, so that
$c_{ij}=h{\cdot}C_{ij}$. In near-isotropic materials like tubulin,
however, $\alpha_{\perp}{\approx}\alpha$ and
$c_{ij}=h{\cdot}C_{ij}(1-\alpha^2)$.

\subsection{Vibrational Modes of empty Nanotubes}

We now consider an infinitely thin 2D plate of finite width
$2{\pi}R$ and an areal mass density $\rho_{2D}$ rolled up to a
nanotube of radius $R$ that is aligned with the $x-$axis. The
linear mass density of the nanotube is related to that of the
plate by
\begin{equation}
\rho_{1D} = 2{\pi}R\rho_{2D} \,.%
\label{Eq2}
\end{equation}
In the long-wavelength limit, represented by
$k=(2{\pi}/{\lambda}){\rightarrow}0$, the longitudinal acoustic
mode of a tubular structure, depicted in Fig.~\ref{fig1}(a),
resembles the stretching mode of a 2D plate~\cite{DT255}. As
mentioned above, the equivalent plate we consider here is a strip
of finite width that is reduced during stretching due to the
nonzero in-plane Poisson ratio $\alpha$.

In the following, we illustrate our computational approach for a
tubular structure by focussing on its longitudinal acoustic mode.
Our derivation, which is described in more detail in Appendices A
and B, starts with the Lagrange function density
\begin{eqnarray}
\label{Eq3}
&&\mathcal{L}\left(\frac{du_x}{dx},\frac{du_x}{dt},x,t\right)
= T - U \\ %
&&\quad = \frac{1}{2} \left[%
        {\rho_{2D}} \left( \frac{du_x}{dt}\right)^2%
      - c_{11} \left(1-\alpha^2\right)
       \left( \frac{du_x}{dx}\right)^2 \right] 2{\pi}R \nonumber \\%
&&\quad = \frac{1}{2} \left[%
           \rho_{1D} \left( \frac{du_x}{dt}\right)^2%
            - c_{LA} \left( \frac{du_x}{dx}\right)^2 \right] \,, %
             \nonumber %
\end{eqnarray}
where
\begin{equation}
c_{LA} = c_{11} \left(1-\alpha^2\right) 2{\pi}R %
      \label{Eq4} %
\end{equation}
is the longitudinal force constant of a 1D nanowire equivalent to
the tubule, and the relation between $\rho_{1D}$ and $\rho_{2D}$
is defined in Eq.~(\ref{Eq2}). The resulting Euler-Lagrange
equation is
\begin{equation}
    \frac{d}{dt} \left(%
        \frac{{\partial}\mathcal{L}}{{\partial}\frac{du_x}{dt}}
                \right)%
 +  \frac{d}{dx} \left(%
        \frac{{\partial}\mathcal{L}}{{\partial}\frac{du_x}{dx}}
                \right)%
    = 0 \,.%
\label{Eq5}
\end{equation}
Using the ansatz
\begin{equation}
u_x = u_{x,0} e^{i(kx-{\omega}t)}%
\label{Eq6}
\end{equation}
we obtain the vibration frequency of the longitudinal acoustic
(LA) mode of the nanotube or nanowire from
\begin{equation}
\omega_{LA} = \sqrt{\frac{c_{11}(1-\alpha^2)}{\rho_{2D}}}~k
            = \sqrt{\frac{c_{LA}}{\rho_{1D}}}~k \,.%
\label{Eq7}
\end{equation}
The prefactor of the crystal momentum $k$ is the longitudinal
speed of sound. As already noted in Ref.~[\onlinecite{Lawler}],
the frequency of the LA mode is independent of the nanotube
radius.

The torsional mode, depicted in Fig.~\ref{fig1}(b), is very
similar to the shear mode of a plate. Consequently, as shown in
Appendix B, the vibration frequency of the torsional acoustic (TA)
mode of the nanotube and the transverse acoustic mode of the plate
should be the same. With $c_{66}$ describing the resistance of the
equivalent plate to shear, we obtain
\begin{equation}
\omega_{TA} = \sqrt{\frac{c_{66}}{\rho_{2D}}}~k \,.%
\label{Eq8}
\end{equation}
Again, prefactor of the crystal momentum $k$ is the corresponding
speed of sound. Similar to the LA mode, the frequency of the TA
mode is independent of the nanotube radius~\cite{Lawler}.

The doubly degenerate flexural acoustic (ZA) mode, depicted in
Fig.~\ref{fig1}(c), differs significantly from the corresponding
bending mode of a plate~\cite{DT255} that involves the plate's
flexural rigidity $D$. The continuum elasticity treatment of the
bending deformation, described in Appendices A and B, leads to
\begin{eqnarray}
\omega_{ZA} &=& %
\sqrt{\frac{{\pi}R^3c_{11}}{\rho_{1D}} %
\left(1+\frac{D}{c_{11}R^2}\right)}~k^2 =
\sqrt{\frac{D_b}{\rho_{1D}}}~k^2 \nonumber \\%
&=& c_{ZA}(R)~k^2\,.%
\label{Eq9}
\end{eqnarray}
Here, $c_{ZA}$ is the effective bending force constant and $D_b$
is the effective beam rigidity of a corresponding nanowire,
defined in Eq.~(\ref{EqA17}).

Finally, the radial breathing mode (RBM) of the nanotube, depicted
Fig.~\ref{fig1}(d), has a nearly $k-$independent frequency given
by~\cite{DT255}
\begin{equation}
\omega_{RBM} = \frac{1}{R}\sqrt{\frac{c_{11}}{\rho_{2D}}}\,.%
\label{Eq10}
\end{equation}
The four vibration modes described above and their functional
dependence on the momentum $k$ and radius $R$ have been partially
described before using an elastic cylindrical shell
model~\cite{{SirenkoPRE96},{WangMicrotubules}}. The schematic
dependence of the vibration frequencies of these modes on $k$ is
shown in Fig.~\ref{fig1}(e). The main expressions for the
vibration frequencies of both 2D and tubular 1D structures are
summarized in Table~\ref{table1}.


\subsection{Vibrational Modes of Liquid-Filled Nanotubes}

We next consider the nanotubes completely filled with a
compressible, but viscosity-free liquid that
may slide without resistance along the nanotube
wall~\cite{Bocquet16}. Since the nanotubes remain straight and
essentially maintain their radius during stretching and torsion,
the frequency $\tilde\omega$ of the LA and TA modes is not
affected by the liquid inside, which remains immobile during the
vibrations. We thus obtain
\begin{equation}
\tilde\omega_{LA}(k) {\approx} \omega_{LA}(k)%
\label{Eq11}
\end{equation}
and
\begin{equation}
\tilde\omega_{TA}(k) = \omega_{TA}(k)\,,%
\label{Eq12}
\end{equation}
where the tilde refers to filling by a liquid.

The only effect of filling by a liquid on the flexural modes is an
increase in the linear mass density to
\begin{equation}
\tilde\rho_{1D}  = \rho_{1D} + {\pi}R^2\rho_l\,,%
\label{Eq13}
\end{equation}
where $\rho_l$ denotes the gravimetric density of the liquid. In
comparison to an empty tube, described by Eq.~(\ref{Eq9}), we
observe a softening of the flexural vibration frequency to
\begin{eqnarray}
\tilde\omega_{ZA} &=& %
\sqrt{\frac{{\pi}R^3c_{11}%
}{\tilde\rho_{1D}}%
\left(1+\frac{D}{c_{11}R^2}\right)}~k^2%
=\sqrt{\frac{D_b}{\tilde\rho_{1D}}}~k^2 \nonumber \\
&=&\tilde{c}_{ZA}(R)~k^2\,.%
\label{Eq14}
\end{eqnarray}
Finally, as we expand in Appendix C, the effect of the contained
liquid on the RBM frequency will depend on the stiffness of the
tubular container. For stiff carbon nanotubes, the RBM mode is
nearly unaffected, whereas the presence of an incompressible
liquid increases $\tilde\omega_{RBM}$ in soft tubules. Thus,
\begin{equation}
\tilde\omega_{RBM} {\agt} \omega_{RBM}\,.%
\label{Eq15}
\end{equation}
The schematic dependence of the four vibration modes on the
momentum $k$ in liquid-filled nanotubes is shown in
Fig.~\ref{fig1}(e). The main expressions for the vibration
frequencies of liquid-filled tubular 1D structures are summarized
in Table~\ref{table1}.

\begin{table}[b]
\caption{\label{table1}%
Summary of expressions derived for the vibrational frequencies
$\omega$ of 1D tubules and 2D plates. $\tilde\omega$ denotes the
frequency of liquid-filled tubules. Equation numbers refer to
the present publication.}%
\begin{ruledtabular}
\begin{tabular}{llll}
   \textrm{Mode} %
 & \quad\textrm{1D Tubules} %
 & \textrm{Equation} %
 & \quad\textrm{2D Plates\footnote{Reference
   [\protect\onlinecite{DT255}].}} \\%
\colrule %
\multirow{3}{*}%
{LA} %
& $\omega_{LA} = \sqrt{\frac{c_{11}(1-\alpha^2)}{\rho_{2D}}}~k$
& Eq.~(\ref{Eq7}) %
& $\omega_{LA} = \sqrt{\frac{c_{11}}{\rho_{2D}}}~k$ \\
{ } %
& $\omega_{LA} = \sqrt{\frac{c_{LA}}{\rho_{1D}}}~k$
& Eq.~(\ref{Eq7}) %
& { } \\
{ } %
& {$\tilde\omega_{LA} = \omega_{LA}$} %
& { } %
& { } \\
\colrule %
\multirow{2}{*}%
{TA} %
& $\omega_{TA} = \sqrt{\frac{c_{66}}{\rho_{2D}}}~k$ %
& Eq.~(\ref{Eq8})%
& $\omega_{TA} = \sqrt{\frac{c_{66}}{\rho_{2D}}}~k$ \\
{ } %
& $\tilde\omega_{TA} = \omega_{TA}$ %
& { } %
& { } \\
\colrule %
\multirow{7}{*}%
{ZA} 
& $\omega_{ZA} =$%
& { } %
& { } \\
{ } 
& $\quad\sqrt{\frac{{\pi}R^3 c_{11}}{\rho_{1D}}
  \left(1+\frac{D}{c_{11}R^2}\right)}~k^2$%
& Eq.~(\ref{Eq9})%
& $\omega_{ZA} = \sqrt{\frac{D}{\rho_{2D}}}$ \\
{ } 
& $\omega_{ZA} = \sqrt{\frac{D_b}{\rho_{1D}}}~k^2 $%
& Eq.~(\ref{Eq9})%
& { } \\
{ } %
& $\tilde\omega_{ZA} =$ %
& { } %
& {} \\
{ } %
& $\quad\sqrt{\frac{{\pi}R^3 c_{11}}{\tilde\rho_{1D}}
  \left(1+\frac{D}{c_{11}R^2}\right)}~k^2$%
& {Eq.~(\ref{Eq14})} %
& { } \\
{ } 
& $\tilde\omega_{ZA} = \sqrt{\frac{D_b}{\tilde\rho_{1D}}}~k^2 $%
& Eq.~(\ref{Eq14})%
& { } \\
\colrule %
\multirow{2}{*}%
{RBM} %
& $\omega_{RBM} = \frac{1}{R}\sqrt{\frac{c_{11}}{\rho_{2D}}}$%
& Eq.~(\ref{Eq10}) %
& {} \\
{ } %
& $\tilde\omega_{RBM} {\agt} \omega_{RBM}$ %
& Eq.~(\ref{Eq15}) 
& {} \\
\end{tabular}
\end{ruledtabular}
\end{table}

\section{Vibrational Modes of Nanotubes in a Surrounding Liquid}

From among the four long-wavelength vibrational modes of nanotubes
illustrated in the left panels of Fig.~\ref{fig1}, the stretching
and the torsional modes are not affected by the presence of a
liquid surrounding the nanotube. We expect the radial breathing
mode in Fig.~\ref{fig1}(d) to couple weakly and be softened by a
small amount in the immersing liquid. The most important effect of
the surrounding liquid is expected to occur for the flexural mode
shown in Fig.~\ref{fig1}(c).

The following arguments and expressions have been developed
primarily to accommodate soft biological structures such as
tubulin-based microtubules, which require an aqueous environment
for their function. We will describe the surrounding liquid by its
gravimetric density $\rho_l$ and viscosity $\eta$. As suggested
above, we will focus our concern on the flexural long-wavelength
vibrations of such structures.

As we will show later on, the flexural modes of idealized,
free-standing biological microtubules are extremely soft. In that
case, the velocity of transverse vibrations will also be very
small and definitely lower than the speed of sound in the
surrounding liquid. Under these conditions, the motion of the
rod-like tubular structure will only couple to the evanescent
sound waves in the surrounding liquid and there will be no
radiation causing damping. The main effect of the immersion in the
liquid will be to increase the effective inertia of the rod. We
may assume that the linear mass density $\rho_{1D}$ of the tubule
in vacuum may increase to $\tilde\rho_{1D}=\rho_{1D}+
{\Delta}\rho_{1D}$ in the surrounding liquid. We can estimate
${\Delta}\rho_{1D}={\Delta}A\rho_l$, where ${\Delta}A$ describes
the increase in the effective cross-section area of the tubule due
to the surrounding liquid that is dragged along during vibrations.
We expect ${\Delta}A{\lesssim}{\pi}R^2$, where $R$ is the radius
of the tubule. The softening of the flexural mode frequency
$\tilde\omega_{ZA}$ due to the increase in $\rho_{1D}$ is
described in Eq.~(\ref{Eq14}).

Next we consider the effect of viscosity of the surrounding liquid
on long-wavelength vibrations of a tubular structure that will
resemble a rigid rod for $k{\rightarrow}0$. Since -- due to
Stoke's paradox -- there is no closed expression for the drag
force acting on a rod moving through a viscous liquid, we will
approximate the rod by a rigid chain of spheres of the same
radius, which are coupled to a rigid substrate by a spring. The
motion for a rigid chain of spheres is the same as of a single
sphere, which is damped by the drag force $F=6{\pi}{\eta}Rv$
according to Stoke's law, where $v$ is the velocity.

The damped harmonic motion of a sphere of radius $R$ and mass $m$
is described by
\begin{equation}
m \frac{d^2{u}}{dt^2} = %
-m{\omega_0^2}u %
-6{\pi}{\eta}R\frac{du}{dt} \,. %
\label{Eq16}
\end{equation}
With the ansatz $u(t)=u_{0}e^{i{\omega}t}$, we get
\begin{equation}
-m\omega^2=%
-m \omega_{0}^2%
-i{\omega}6{\pi}{\eta}R %
\label{Eq17}
\end{equation}
and thus
\begin{equation}
\omega = %
\pm\sqrt{\omega_{0}^2-\left(\frac{3{\pi}{\eta}R}{m}\right)^2}%
+ i\frac{3{\pi}{\eta}R}{m} \,. %
\label{Eq18}
\end{equation}
Assuming that the damping is small, we can estimate the energy
loss described by the $Q-$factor
\begin{equation}
Q = \omega_{0} \frac{m}{3{\pi}{\eta}R} %
= \frac{2}{3} \frac{m}{R} \frac{f_0}{\eta} \,, %
\label{Eq19}
\end{equation}
where $f_0=\omega_0/(2{\pi})$ is the harmonic vibration frequency.
In a rigid string of masses separated by the distance $2R$, the
linear mass density is related to the individual masses by
$\rho_{1D}=m/(2R)$. Then, the estimated value of the $Q-$factor
will be
\begin{equation}
Q=\frac{4}{3}\rho_{1D}\frac{f_0}{\eta} \,. %
\label{Eq20}
\end{equation}

%
%
%
%
%
%
%
%
%
%

\section{Computational Approach to Determine the Elastic
         Response of Carbon Nanotubes}

We determine the elastic response and elastic constants of an
atomically thin graphene monolayer, the constituent of CNTs, using
{\em ab initio} density functional theory (DFT) as implemented in
the {\textsc SIESTA}~\cite{SIESTA} code. We use the
Perdew-Burke-Ernzerhof (PBE)~\cite {PBE} exchange-correlation
functional, norm-conserving Troullier-Martins
pseudopotentials~\cite{Troullier91}, and a double-$\zeta$ basis
including polarization orbitals. To determine the energy cost
associated with in-plane distortions, we sampled the Brillouin
zone of a 3D superlattice of non-interacting layers by a
$20{\times}20{\times}1$ $k$-point grid~\cite{Monkhorst-Pack76}. We
used a mesh cutoff energy of $180$~Ry and an energy shift of
10~meV in our self-consistent total energy calculations, which has
provided us with a precision in the total energy of
${\leq}2$~meV/atom. The same static approach can be applied to
other layered materials that form tubular structures.

\begin{figure}[t]
\includegraphics[width=1.0\columnwidth]{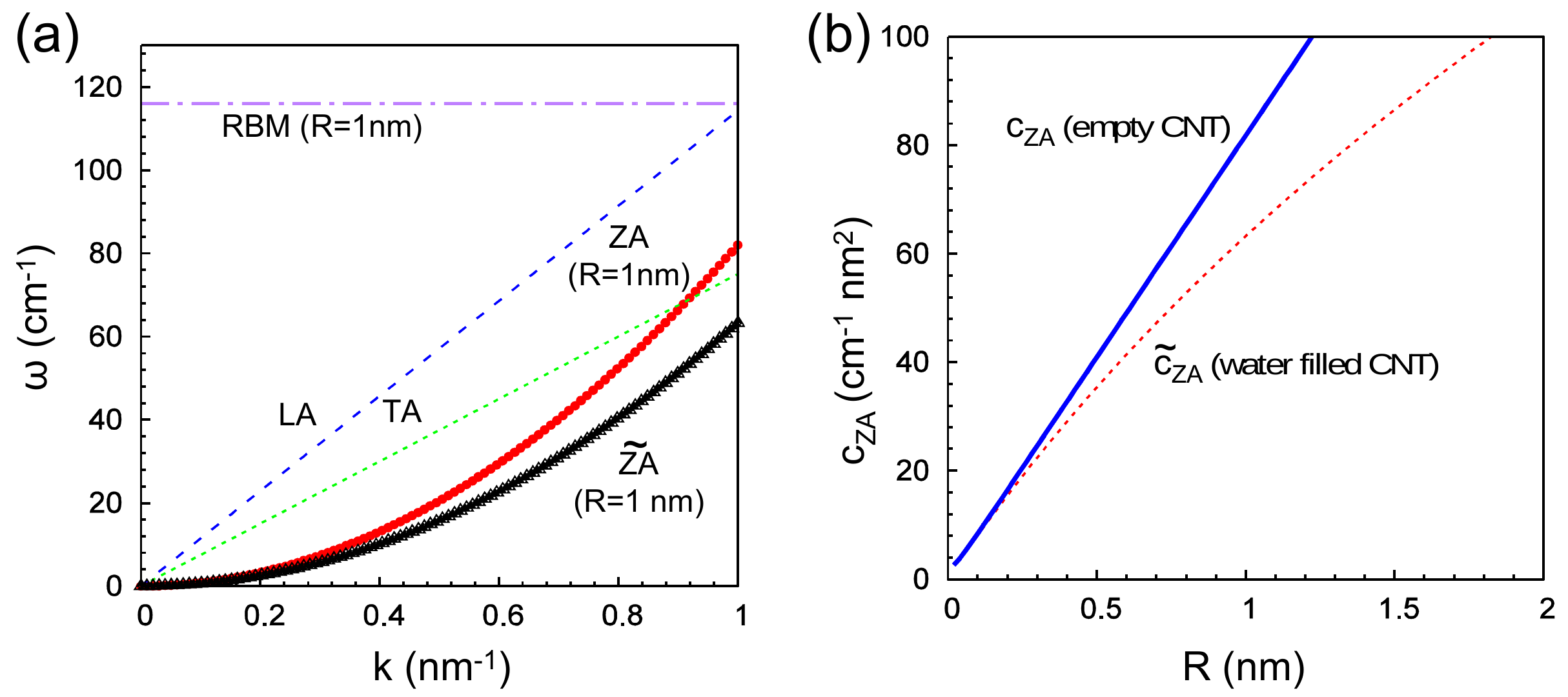}
\caption{(Color online) (a) Frequency of vibrational modes
depicted in Fig.~\protect\ref{fig1}(a) in empty and water-filled
carbon nanotubes. (b) Dependence of the flexural coefficient
$c_{ZA}(R)$, defined in Eq.~(\protect\ref{Eq9}), on the radius $R$
of empty and water-filled carbon nanotubes. The tilde denotes
liquid-filled nanotubes.%
\label{fig2} }
\end{figure}

\section{Results}

To illustrate the usefulness of our approach for all tubular
structures, we selected two extreme examples. Nanometer-wide CNTs
have been characterized well as rigid structures able to support
themselves in vacuum. Tubulin-based microtubules, on the other
hand, are significantly wider and softer than carbon nanotubes.
These biological structures require an aqueous environment for
their function.


\subsection{Carbon Nanotubes}

\begin{table}[b]
\caption{\label{table2}%
Elastic behavior of a 2D graphene monolayer. $c_{11}$, $c_{22}$,
$c_{66}$ are 2D elastic stiffness constants defined in
Eq.~(\ref{Eq1}). $\alpha$ is the in-plane Poisson ratio, $D$ is
the flexural rigidity, and $\rho_{2D}$ is the areal mass density.
These values can be used directly to calculate long-wavelength
acoustic frequencies $\omega(k)$ using the expressions in
Table~\ref{table1} and the speed of sound $v_{LA}$ and $v_{TA}$.
Present values are compared to published data.
}%
\begin{ruledtabular}
\begin{tabular}{lcr}
Quantity & Present result & Literature values \\
\colrule %
$c_{11}=c_{22}$ %
& 352.6~N/m  %
& 342~N/m~\footnote{Reference [\protect\onlinecite{Politano}].}\\
\colrule %
$c_{66}$ %
& 146.5~N/m  %
& 144~N/m~$^{\rm a}$\\
\colrule %
$\alpha$ %
& 0.17   %
& 0.19~$^{\rm a}$\\
\colrule %
$D$ %
& $0.22$~GPa${\cdot}$nm$^3$ %
& $0.225$~GPa${\cdot}$nm$^3$~\footnote{Reference [\protect\onlinecite{Lu2009}].}\\
\colrule %
\multirow{2}{*}%
{$R~\omega_{RBM}=\sqrt{c_{11}/\rho_{2D}}$} %
& \multirow{2}{*}{$116$~cm$^{-1}$nm} %
& $116$~cm$^{-1}$nm~\footnote{Reference [\protect\onlinecite{sanchez1999}].}\\
%
{ } %
& { } %
& $108$~cm$^{-1}$nm~\footnote{Reference [\protect\onlinecite{maultzsch2005}].}\\
\colrule %
\multirow{2}{*}%
{$v_{LA}$} %
& \multirow{2}{*}{21.5~km/s} %
& 22~km/s~~$^{\rm a}$\\
{ } %
& { } %
& ${\approx}21$~km/s~\footnote{Reference [\protect\onlinecite{Lawler}].}\\
\colrule %
$v_{TA}$ %
& 14.1~km/s %
& 14~km/s~$^{\rm a,b}$\\
\end{tabular}
\end{ruledtabular}
\end{table}

The elastic behavior of carbon nanotubes can be described using
quantities previously obtained using DFT calculations for
graphene~\cite{DT255}. The calculated elements of the elastic
stiffness matrix (\ref{Eq1}) are $c_{11}=c_{22}=352.6$~N/m,
$c_{12}=59.6$~N/m, and $c_{66}= 146.5 N/m$, all in very good
agreement with experimental results~\cite{Politano}. The
calculated in-plane Poisson ratio $\alpha=c_{12}/c_{11}=0.17$ is
also close to the experimentally estimated value for
graphene~\cite{Politano} of $\alpha_{expt}=0.19$.
The calculated flexural rigidity of a graphene plate is
$D=0.22$~GPa$\cdot$nm$^3$.
The calculated 2D mass density of graphene
$\rho_{2D}=0.743{\cdot}10^{-6}$~kg/m$^2$ translates to
$\rho_{1D}=0.743{\cdot}10^{-6}$~kg/m$^2$${\cdot}2{\pi}R$ for
nanotubes of radius $R$.

The phonon dispersion relations ${\omega}(k)$ depend primarily on
the radius and not the specific chiral index $(n,m)$ of carbon
nanotubes and are presented in Fig.~\ref{fig2}(a) for the
different polarizations. The LA and TA mode frequencies are almost
independent of the nanotube radius for a given $k$. The
corresponding group velocities at $k{\to}0$ give the longitudinal
speed of sound of $v_{LA}=d{\omega}_{LA}/dk=21.5$~km/s and the
speed of sound with torsional polarization of
$v_{TA}=d{\omega}_{TA}/dk=14.1$~km/s.

The flexural or bending ZA mode does depend on the nanotube radius
through the proportionality constant $c_{ZA}(R)$, defined in
Eq.~(\ref{Eq9}), which is plotted as a function of $R$ in
Fig.~\ref{fig2}(b). The dispersion of the ZA mode in a CNT of
radius $R=1$~nm is shown in Fig.~\ref{fig2}(a). Also the RBM
frequency depends on the nanotube radius according to
Eq.~(\ref{Eq10}). We find the value
$\sqrt{c_{11}/\rho_{2D}}=116$~cm$^{-1}$nm of the prefactor of
$R^{-1}$ in Eq.~(\ref{Eq10}) to agree well with the published
theoretical value~\cite{sanchez1999} of $116$~cm$^{-1}$nm and with
the value of $108$~cm$^{-1}$nm, obtained by fitting a set of
observed Raman frequencies~\cite{maultzsch2005}. The calculated
value ${\omega}_{RBM}=116$~cm$^{-1}$ for CNTs with $R=1$~nm is
shown in Fig.~\ref{fig2}(a).

Filling the CNT with a liquid of density $\rho_l$ increases its
linear density $\rho_{1D}$ according to Eq.~(\ref{Eq13}). For a
nanotube filled with water of density $\rho_l=1$~g/cm$^3$, the
radius-dependent quantity $\tilde{c}_{ZA}(R)$, defined in
Eq.~(\ref{Eq14}), is plotted as a function of $R$ in
Fig.~\ref{fig2}(b). The dispersion of the $\tilde{\rm{ZA}}$ mode
in a water-filled CNT of radius $R=1$~nm is shown in
Fig.~\ref{fig2}(a).

Elastic constants calculated in this work, and results derived
using the present continuum elasticity approach are listed and
compared to literature data in Table~\ref{table2}.

\begin{figure}[t]
\includegraphics[width=1.0\columnwidth]{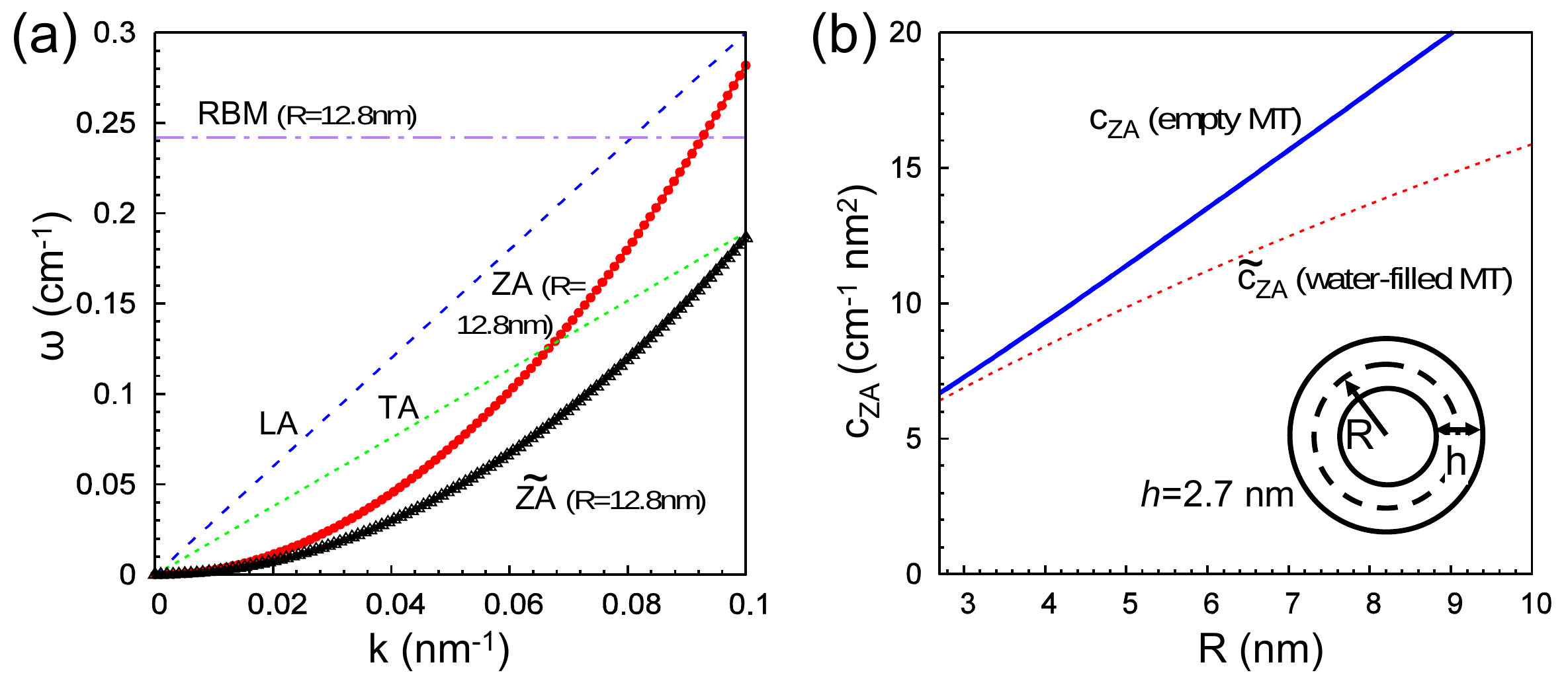}
\caption{(Color online) (a) Frequency of vibrational modes
depicted in Fig.~\protect\ref{fig1}(a) in empty and water-filled
tubulin-based microtubules. (b) Dependence of the flexural
coefficient $c_{ZA}(R)$, defined in Eq.~(\protect\ref{Eq9}), on
the radius $R$ of empty and water-filled tubulin-based
microtubules. The tilde denotes liquid-filled tubules.%
\label{fig3} }
\end{figure}

\subsection{Tubulin-Based Microtubules}

To describe phonon modes in tubulin-based microtubules, we depend
on published experimental data~\cite{venier1994analysis} for
microtubules with an average radius $R=12.8$~nm and a wall
thickness $h=2.7$~nm.
The observed density of the tubule wall material
${\rho}=1.47$~g/cm$^3$ translates to
${\rho}_{2D}=4.0{\cdot}10^{-6}$~kg/m$^2$. The estimated Young's
modulus of the wall material is $E=0.5$~GPa and the flexural beam
rigidity of these microtubules with $R=12.8$~nm is
$D_b=0.9{\cdot}10^{-23}$~$N{\cdot}m^2$. We may use the
relationship between $D_b$ and $D$, defined in Eq.~(\ref{EqA16}),
to map these values onto the elastic 2D wall material and obtain
$c_{11}=E{\cdot}h=1.4$~N/m and $D=2.71$~GPa$\cdot$nm$^3$.
In a rough approximation, tubulin can be considered to be
isotropic, with a Poisson ratio ${\alpha}=0.25$. Further assuming
that $c_{11}=c_{22}$, we estimate
$c_{66}=(c_{11}-c_{12})/2=c_{11}(1-{\alpha})/2=0.5$~N/m.

One fundamental difference between tubulin-based microtubules and
systems such as CNTs is that the former necessitate an aqueous
environment for their shape and function. Thus, the correct
description of microtubule deformations and vibrations requires
addressing the complete microtubule-liquid system, which would
exceed the scope of this study. We rather resorted to the
expressions derived in the Subsection on nanotubes in a
surrounding liquid to at least estimate their $Q-$factor in an
aqueous environment. We used %
$\rho_{1D}=7{\times}10^{-13}$~kg/m for a water-filled microtubule, %
$\eta=10^{-3}$~Pa$\cdot$s and %
$f_0=10^9$~Hz, which provided us with the value $Q{\approx}1.2$.
In other words, flexural vibrations in microtubules should be
highly damped in aqueous environment, so their frequency should be
very hard to measure. Consequently, the only available comparison
between our calculations and experimental data should be made for
static measurements.

An elegant way to experimentally determine the effective beam
rigidity of individual tubulin-based microtubules involves the
measurement of buckling caused by applying an axial buckling force
using optical tweezers. Experimental results for the effective
beam rigidity have been obtained in this way for microtubules that
are free of the paclitaxel stabilizing agent and contain 14
tubulin protofilaments, which translates to the effective radius
$R{\approx}9.75$~nm. The observed values range from
$D_b=3.7{\pm}0.8{\times}10^{-24}$~Nm$^2$~\cite{Felgner509} to
$D_b=7.9{\pm}0.7{\times}10^{-24}$~Nm$^2$~\cite{kikumoto2006}, in
good agreement with our calculated value
$D_b=4.2{\times}10^{-24}$~Nm$^2$. Our estimated value
$D_b=6.2{\times}10^{-24}$~Nm$^2$ for wider tubules with 16
protofilaments is 49\% larger than for the narrower tubules with
14 protofilaments. The corresponding increase by 49\% has been
confirmed in a corresponding experiment~\cite{kikumoto2006}.

Next, we may still use the oversimplifying assumption that tubulin
microtubules may exist in the vacuum and could be described by the
above-derived continuum values~\cite{WangMicrotubules}. In this
way, we may compare our results to published theoretical results.
The calculated phonon dispersion relations ${\omega}(k)$ for most
common microtubules with the radius $R=12.8$~nm are presented in
Fig.~\ref{fig3}(a). The LA and TA mode frequencies are independent
of the tubule radius. From their slope, we get the longitudinal
speed of sound $v_{LA}=d{\omega}_{LA}/dk=0.56$~km/s and the speed
of sound with torsional polarization
$v_{TA}=d{\omega}_{TA}/dk=0.36$~km/s. For the sake of comparison,
we extracted the $v_{LA}$ value based on the elastic cylindrical
shells model with $E=2.0$~GPa from Ref.~\cite{WangMicrotubules}.
Extrapolating to the value $E=0.5$~GPa in our set of parameters
using the relationship $v_{LA}{\propto}\sqrt{E}$, we obtained
$v_{LA}=0.59$~km/s, in excellent agreement with our calculated
value.

The flexural or bending ZA mode depends on the tubule radius
through the proportionality constant $c_{ZA}(R)$, defined in
Eq.~(\ref{Eq9}), which is plotted as a function of $R$ in
Fig.~\ref{fig3}(b). The dispersion of the ZA mode in a microtubule
of radius $R=12.8$~nm is shown in Fig.~\ref{fig3}(a).

Also the RBM frequency depends on the nanotube radius according to
Eq.~(\ref{Eq10}). For $R=12.8$~nm, we obtain
${\omega}_{RBM}=0.24$~cm$^{-1}$, as seen in Fig.~\ref{fig3}(a).

To describe the increase in the linear density $\rho_{1D}$ of a
microtubule filled with a liquid of density $\rho_l$, we have to
account for the finite thickness $h$ of the wall and replace the
radius $R$ by $R-h/2$ in Eq.~(\ref{Eq13}). Considering water of
density $\rho_l=1$~g/cm$^3$ as the filling medium, we plot the
radius-dependent quantity $\tilde{c}_{ZA}(R)$, defined in
Eq.~(\ref{Eq14}), as a function of $R$ in Fig.~\ref{fig3}(b). The
dispersion of the $\tilde{\rm{ZA}}$ mode in a water-filled CNT of
radius $R=12.8$~nm is shown in Fig.~\ref{fig3}(a).

Our results in Fig.~\ref{fig3}(a) suggest soft vibration in the
GHz range, in agreement with other theoretical
estimates~\cite{{SirenkoPRE96},{WangMicrotubules}}. As mentioned
above, all these vibrations will be hight damped in an aqueous
environment to the low $Q-$factor.


\section{Discussion}

Our study has been motivated by the fact that the conventional
approach to calculate the frequency spectrum, based on an
atomistic calculation of the force-constant matrix, does not
provide accurate frequencies for long-wavelength soft acoustic
modes in quasi-1D tubular structures. We should note that the
atomistic approach is quite adequate to determine frequencies of
the optical and of short-wavelength acoustic modes. But for
long-wavelength acoustic modes, the excessive demand on supercell
size and basis convergence often yields imaginary vibration
frequencies as an artifact of insufficient convergence.

As a viable alternative to tedious atomistic calculations of the
force-constant matrix of complex tubular systems, we propose here
a continuum elasticity approach to determine the frequency of
long-wavelength acoustic modes in tubular structures that does not
require the thickness of the wall as an input. Our approach for
quasi-1D structures is based on the successful description of
corresponding modes in 2D structures~\cite{DT255}. The continuum
elasticity approach introduced in this study has a significant
advantage over the 3D elastic modulus approach, which which has
lead to inconsistencies in describing the elastic behavior of thin
walls and membranes~\cite{wu1985}. Using this approach, we obtain
for the first time quantitative results for systems ranging from
stiff CNTs to much wider and softer protein microtubules.

We found that the elastic behavior of the wall material can be
determined accurately by static calculations of 2D plate subjected
to small deformation or by elastic measurements. The validity of
predictions based on this approach is limited to long-wavelength
vibrations and large-radius nanotubes, both of which would require
extraordinary computational resources in atomistic calculations.
In particular, the flexural ZA modes with their
${\omega}{\propto}k^2$ momentum dependence are known to be very
hard to reproduce by {\em ab initio} calculations near the
$\Gamma$ point~\cite{Ling10}.

For the sake of completeness, we have also derived the
Euler-Lagrange equations of motion required to describe all
long-wavelength acoustic modes and present the detailed derivation
in the Appendix.

Of course, the frequency of the ZA modes is expected to be much
softer than that of the LA mode in any nanotube or nanowire. Since
${\omega}_{ZA}{\propto}k^2$ whereas ${\omega}_{LA}{\propto}k$,
expressions derived here for the long-wavelength limit would lead
to the unrealistic behavior ${\omega}_{ZA}>{\omega}_{LA}$ for
large values of $k$. This limits the $k-$range, for which our
formalism is valid in the dispersion relations presented in
Figs.~\ref{fig2}(a) and \ref{fig3}(a). In a crystalline tubule,
$k$ is restricted to typically an even smaller range given by the
size of the 1D Brillouin zone.

For systems with a vanishing Poisson ratio $\alpha$, the radial
breathing mode (RBM) should be decoupled from the longitudinal
acoustic or stretching mode. However, as discussed in Appendix D,
most systems have a non-vanishing value of $\alpha$. In that case,
the two modes mix and change their character beyond the wavevector
$k=1/R$, where ${\omega}_{LA}(k){\approx}{\omega}_{RBM}$, as
discussed previously~\cite{WangMicrotubules}. At smaller values of
$k$, coupling between the LA mode and the RBM modifies the
frequency of these modes by only ${\approx}1$\% in CNTs.

Our model allows a simple extension from empty to liquid-filled
nanotubes. We find that presence of a filling liquid does not
affect longitudinal acoustic and torsional acoustic modes to a
significant extent, as shown in Appendix C, but softens the
flexural modes. We also expect the pressure wave of the liquid to
couple to the RBM beyond the wavevector
$k{\approx}{\omega}_{RBM}/v_p$, where $v_p$ is the speed of the
propagating pressure wave.

To demonstrate the universality of our approach, we also
considered microtubules formed of the proteins $\alpha$- and
$\beta$-tubulin. These are responsible for maintaining the shape
and elasticity of cells, but are too complex for an atomistic
description to predict vibration spectra. From a computational
point of view, the necessity to include the aqueous environment in
the description of tubulin-based microtubules adds another layer
of complexity to the problem.

Our basic finding that microtubule motion and vibrations are
overdamped in the natural aqueous environment, with a $Q-$factor
of the order of unity, naturally explains the absence of
experimental data reporting observation of motion, dynamical shape
changes and vibrations in these protein-based systems. Among
static measurements of the elastic behavior of microtubules,
optical tweezers appear to be the optimum way to handle and deform
individual microtubules in order to determine their effective beam
rigidity $D_b$. In this static scenario, we find our description
of the beam rigidity precise enough to compare with experimental
data. The reported dependence of $D_b$ on the cube of the
radius~\cite{kikumoto2006} is reflected in our corresponding
expression for $D_b$ in Eq.~(\ref{EqA16}). In the case of
tubulin-based microtubules, we find the leading term in $D_b$ to
be indeed proportional to $R^3$ and to be much larger than the
second term, which is proportional to $R$.



\section{Summary and Conclusions}

Addressing the shortcoming of conventional atomistic calculations
of long-wavelength acoustic frequencies in tubular structures,
which often yield numerical artifacts, we have developed an
alternative computational approach representing an adaptation of
continuum elasticity theory to 2D and 1D structures. Since 1D
tubular structures can be viewed as 2D plates of finite width
rolled up to a cylinder, we have taken advantage of the
correspondence between 1D and 2D structures to determine their
elastic response to strain. In our approach, computation of
long-wavelength acoustic frequencies does not require the
determination and diagonalization of a large, momentum-dependent
dynamical matrix. Instead, the simple expressions we have derived
for the acoustic frequencies $\omega(k)$ use only four elements of
a $k-$independent 2D elastic matrix, namely $c_{11}$, $c_{22}$,
$c_{12}$, and $c_{66}$, as well as the value of the flexural
rigidity $D$ of the 2D plate constituting the wall. These five
numerical values can easily be obtained using static calculations
for a 2D plate. Even though the scope of our approach is limited
to long-wavelength acoustic modes, we found that the accuracy of
the calculated vibration frequencies surpasses that of
conventional atomistic {\em ab initio} calculations. Starting with
a Lagrange function describing longitudinal, torsional, flexural
and radial deformations of empty or liquid-filled tubular
structures, we have derived corresponding Euler-Lagrange equations
to obtain simple expressions for the vibration frequencies of the
corresponding modes. We have furthermore shown that longitudinal
and flexural acoustic modes of tubules are well described by those
of an elastic beam resembling a nanowire. Using our simple
expressions, we were able to show that a pressure wave in the
liquid contained in a stiff carbon nanotube has little effect on
its RBM frequency, whereas the effect of a contained liquid on the
RBM frequency in much softer tubulin tubules is significant. We
found that presence of water in the native environment of tubulin
microtubules reduces the $Q-$factor to such a degree that flexural
vibrations can hardly be observed. We also showed that the
coupling between long-wavelength LA modes and the RBM can be
neglected. We have found general agreement between our numerical
results for biological microtubules and carbon nanotubes and
available experimental data.


\section*{Acknowledgments}

\begin{acknowledgments}
We acknowledge useful discussions with Jie Guan. A.G.E.
acknowledges financial support by the South African National
Research Foundation. D.L. and D.T. acknowledges financial support
by the NSF/AFOSR EFRI 2-DARE grant number EFMA-1433459.
\end{acknowledgments}


\section{Appendix}

Material in the Appendix provides detailed derivation of
expressions used in the main text and considers specific limiting
cases. In Appendix A, we derive the Lagrange function for
stretching, torsional and bending modes of tubular structure. In
Appendix B, we derive analytical expressions for the frequencies
of the corresponding vibration modes using the Euler-Lagrange
equations. The effect of a liquid contained inside a CNT on its
RBM frequency is investigated in Appendix C. The coupling between
the longitudinal acoustic mode and the RBM in a CNT due to its
non-vanishing Poisson ratio is discussed in Appendix D.

\renewcommand\thesubsection{\Alph{subsection}}
\setcounter{subsection}{0} %
\setcounter{equation}{0} %
\renewcommand{\theequation}{A\arabic{equation}}

\subsection{A. Lagrange Function of a Strained Nanotube}

\subsubsection{\bf\em Stretching}

Let us consider a nanotube of radius $R$ aligned with the $x-$axis
and its response to tensile strain ${du_x}/{dx}$ applied uniformly
along the $x-$direction. The strain energy will be the same as
that of a 2D strip of width $y=2{\pi}R$ lying in the $xy-$plane
that is subject to the same condition.

Assuming that the width of the strip is constrained to be
constant, the strain energy per length is given by
\begin{equation}
U_x = \frac{1}{2}~c_{11} \left( \frac{du_x}{dx}\right)^2 2\pi R \,.%
\label{EqA1}
\end{equation}
For a nonzero Poisson ratio $\alpha$, stretching the strip by
$du_x/dx$ will reduce its width by $du_y/dy={\alpha}du_x/dx$ and
its radius $R$, as shown in Fig.~\ref{fig1}(a). Releasing the
constrained width will release the energy $U_y=-{\alpha}^2U_x$.
The total strain energy per length of a nanotubule or an
equivalent 1D nanowire is the sum $U=U_x+U_y$ and is given by
\begin{eqnarray}
U &=& \frac{1}{2} c_{11} \left(1-\alpha^2\right) %
    \left( \frac{du_x}{dx}\right)^2 2\pi R \nonumber \\ %
    &=& \frac{1}{2} c_{LA} \left( \frac{du_x}{dx}\right)^2 \,.
\label{EqA2}
\end{eqnarray}
Here, $c_{LA}=2{\pi}R c_{11}(1-\alpha^2)$ is the the longitudinal
force constant of a 1D nanowire equivalent to the tubule, defined
in Eq.~(\ref{Eq4}).

In the harmonic regime, we will consider only small strain values.
Releasing the strain will cause a vibration in the $x$-direction
with the velocity $v_x=du_x/dt$. Then, the kinetic energy density
of the strip will be given by
\begin{equation}
T = \frac{1}{2} {\rho_{2D}} \left( \frac{du_x}{dt}\right)^2 2\pi R %
  = \frac{1}{2} {\rho_{1D}} \left( \frac{du_x}{dt}\right)^2\,,%
\label{EqA4}
\end{equation}
where ${\rho_{2D}}$ is the areal mass density of the equivalent
strip that is related to ${\rho_{1D}}$ by Eq.~(\ref{Eq2}). The
Lagrangian density is then given by
\begin{eqnarray}
\label{EqA5}
&&\mathcal{L}\left(\frac{du_x}{dx},\frac{du_x}{dt},x,t\right)
= T - U \\ %
&&\quad = \frac{1}{2} \left[%
        {\rho_{2D}} \left( \frac{du_x}{dt}\right)^2%
      - c_{11} \left(1-\alpha^2\right)
       \left( \frac{du_x}{dx}\right)^2 \right] 2{\pi}R %
             \,, \nonumber \\%
&&\quad = \frac{1}{2} \left[%
           \rho_{1D} \left( \frac{du_x}{dt}\right)^2%
            - c_{LA} \left( \frac{du_x}{dx}\right)^2 \right] \,. %
             \nonumber%
\end{eqnarray}


\subsubsection{\bf\em Torsion}

The derivation of the Euler-Lagrange equation for the torsional
motion is very similar to that for the longitudinal motion. The
main difference is that the displacement $u_{\phi}$ is normal to
the propagation direction $x$. To obtain the corresponding
equations, we need to replace $u_x$ by $u_{\phi}$ and
$c_{11}(1-\alpha^2)$ by $c_{66}$ in Eqs.~({A1})-({A4}). The
Lagrangian density is then given by
\begin{eqnarray}
\label{EqA6}
&&\mathcal{L}%
\left(\frac{du_{\phi}}{dx},\frac{du_{\phi}}{dt},x,t\right)
= T - U \\ \nonumber %
&&\quad = \frac{1}{2} \left[%
          {\rho_{2D}}\left( \frac{du_{\phi}}{dt}\right)^2%
           - c_{66} \left( \frac{du_{\phi}}{dx}\right)^2\right] %
           2{\pi}R%
           \,. \nonumber%
\end{eqnarray}


\subsubsection{\bf\em Bending}

Bending a nanotube of radius $R$ is equivalent to its
transformation to a segment of a nanotorus of radius $R_t$.
Initially, we will assume that $c_{11}=0$ and $D>0$ in the given
nanotorus segment, so the strain energy would contain only an
out-of-plane component. We first consider a straight nanotube of
radius $R$ formed by rolling up a %
plate of width $2{\pi}R$. The corresponding out-of-plane strain
energy per nanotube segment length is
\begin{equation}
U = \frac{1}{2} \frac{D}{R^2} \left( 2{\pi}R\right) %
= \frac{\pi D}{R} \,. %
\label{EqA7}
\end{equation}
The corresponding expression for the total out-of-plane strain
energy in a nanotorus is~\cite{DT245},
\begin{equation}
U = 2 {\pi}^2 D \frac{R_t^2}{R\sqrt{(R_t+R)(R_t-R)}} \,.
\label{EqA8}
\end{equation}
Divided by the average perimeter length $2{\pi}R_t$, we obtain the
out-of-plane energy of the torus per nanotube segment length
\begin{equation}
U = \pi D \frac{R_t}{R\sqrt{(R_t+R)(R_t-R)}} \,.%
\label{EqA9}
\end{equation}
Assuming that the torus radius is much larger than the nanotube
radius, $R_t{\gg}R$, we can Taylor expand $U$ in Eq.~(\ref{EqA9})
and neglect higher-order terms in $(R/R_t)$, which leads to
\begin{equation}
U = \frac{{\pi}D}{R}\left(1+\frac{1}{2}%
\left( \frac{R}{R_t}\right)^2 \right) \,.%
\label{EqA10}
\end{equation}
Comparing the out-of-plane strain energy of a bent nanotube in
Eq.~(\ref{EqA10}) and that of a straight nanotube in
Eq.~(\ref{EqA7}), the change in out-of-plane strain energy per
segment length associated with bending turns out to be
\begin{equation}
U = \frac{1}{2}{\pi}DR \left(\frac{1}{R_t}\right)^2\,. %
\label{EqA11}
\end{equation}
During the flexural or bending vibration mode, the local curvature
$1/R_t=d^2u_z/dx^2$ changes along the tube, yielding the local
in-plane strain energy per nanotube segment length of
\begin{equation}
U = \frac{1}{2}{\pi}DR \left(\frac{d^2u_z}{dx^2}\right)^2\,. %
\label{EqA12}
\end{equation}
Next, we consider the in-plane component of strain, obtained by
assuming $c_{11}>0$ and $D=0$ in a given nanotorus segment. There
is nonzero strain in a nanotube deformed to a very wide torus with
$R_t{\gg}R$ even if its cross-section and radius $R$ were not to
change in this process. The reason is that the %
wall of the nanotube undergoes stretching along the outer and
compression along the inner torus perimeter in this process. This
amounts to a total in-plane strain energy~\cite{DT245}
\begin{equation}
U=\frac{{\pi}^2 c_{11} R^3}{R_t} %
\label{EqA13}
\end{equation}
for the entire torus with an average perimeter of $2{\pi}R_t$ in
relation to a straight nanotube of length $2{\pi}R_t$. Thus, the
in-plane strain energy within the torus per segment length is
\begin{equation}
U = \frac{1}{2}{\pi}c_{11}R^3\left(\frac{1}{R_t}\right)^2\,. %
\label{EqA14}
\end{equation}
Considering local changes in curvature $1/R_t=d^2u_z/dx^2$ during
the bending vibrations of a nanotube, the local in-plane strain
energy per nanotube segment length becomes
\begin{equation}
U=\frac{1}{2}{\pi}c_{11}R^3 \left(\frac{d^2 u_z}{dx^2}\right)^2 \,. %
\label{EqA15}
\end{equation}
The strain energy in the deformed nanotube per length is the sum
of the in-plane strain energy in Eq.~(\ref{EqA15}) and the
out-of-plane strain energy in Eq.~(\ref{EqA12}), yielding
\begin{equation}
U =%
\frac{1}{2} ({\pi}c_{11}R^3 + {\pi}DR)%
\left(\!\frac{d^2{u_z}}{dx^2}\!\right)^2\!=
\frac{1}{2} D_b %
\left(\!\frac{d^2{u_z}}{dx^2}\!\right)^2\!,%
\label{EqA16}
\end{equation}
where
\begin{equation}
D_b ={\pi}c_{11}R^3 + {\pi}DR%
\label{EqA17}
\end{equation}
is the effective beam rigidity of a corresponding nanowire. The
kinetic energy of a bending nanotube or nanowire segment is given
by
\begin{equation}
T=\frac{1}{2}\rho_{1D} \left(\frac{du_z}{dt}\right)^2 \,. %
\label{EqA18}
\end{equation}
This leads to the Lagrangian density
\begin{eqnarray}
\label{EqA19}
&&\mathcal{L}\left(\frac{d^2u_z}{dx^2},\frac{du_z}{dt},x,t\right)
= T - U \\ \nonumber %
&&\quad= \frac{1}{2} \left[%
        {\rho_{1D}} \left(\!\frac{du_z}{dt}\!\right)^2 %
        \mkern-10mu 
      - {{\pi}c_{11}R^3}\left(\!1+\frac{D}{c_{11}R^2}\!\right)\!\!%
        \left(\!\frac{d^2u_z}{dx^2}\!\right)^2%
               \right]\,. \nonumber%
\end{eqnarray}
%

\subsection{B. Derivation of Euler-Lagrange Equations of Motion
 for Deformations of a Nanotube using Hamilton's Principle}

\subsubsection{\bf\em Stretching}

The Euler-Lagrange equation for stretching a tube or a plate
is~\cite{DT255}
\begin{equation}
    \frac{d}{dt} \left(%
        \frac{{\partial}\mathcal{L}}{{\partial}\frac{du_x}{dt}}
                \right)%
 +  \frac{d}{dx} \left(%
        \frac{{\partial}\mathcal{L}}{{\partial}\frac{du_x}{dx}}
                \right)%
    = 0 \,.%
\label{EqA20}
\end{equation}
Inserting the Lagrangian of Eq.~(\ref{EqA5}) in the Euler-Lagrange
Eq.~(\ref{EqA20}) yields the wave equation for longitudinal
vibrations of the tubule or the equivalent nanowire
\begin{eqnarray}
 2\pi R{\rho_{2D}}\frac{d^2u_x}{dt^2} %
 - 2 \pi R c_{11}\left(1-\alpha^2\right) %
 \frac{d^2u_x}{dx^2} &=& \nonumber \\
       {\rho_{1D}}\frac{d^2u_x}{dt^2} %
 - c_{LA} \frac{d^2u_x}{dx^2} &=& 0 \,.
\label{EqA21}
\end{eqnarray}
The nanotube radius $R$ drops out and we obtain
\begin{eqnarray}
{\rho_{2D}}\frac{d^2u_x}{dt^2} %
- c_{11}\left(1-\alpha ^2 \right) %
\frac{d^2u_x}{dx^2} &=& \nonumber \\%
       {\rho_{1D}}\frac{d^2u_x}{dt^2} %
 - c_{LA} \frac{d^2u_x}{dx^2} &=& 0 \,.
\label{EqA22}
\end{eqnarray}
This wave equation can be solved using the ansatz
\begin{equation}
u_x = u_{x,0} e^{i(kx-{\omega}t)}%
\label{EqA23}
\end{equation}
to yield
\begin{equation}
{\rho_{2D}}{\omega}^2 = c_{11}
\left(1-\alpha^2 \right) k^2 %
\label{EqA24}
\end{equation}
for a tubular structure or
\begin{equation}
{\rho_{1D}}{\omega}^2 = c_{LA} k^2 %
\label{EqA25}
\end{equation}
for an equivalent 1D nanowire. This finally translates to the
desired form
\begin{equation}
\omega_{LA} = \sqrt{\frac{c_{LA}}{\rho_{1D}}}~k = %
              \sqrt{\frac{c_{11}(1-\alpha^2)}{\rho_{2D}}}~k \,.%
\label{EqA26}
\end{equation}
which is identical to Eq.~(\ref{Eq7}).


\subsubsection{\bf\em Torsion}

The Lagrangian $\mathcal{L}(du_{\phi}/dx,du_{\phi}/dt,x,t)$ in
Eq.~(\ref{EqA6}), which describes the torsion of a tubule, has a
similar form as the Lagrangian in Eq.~(\ref{EqA5}). To obtain the
equations for torsional motion from those for stretching motion,
we need to replace $u_x$ by $u_{\phi}$ and $c_{11}(1-\alpha^2)$ by
$c_{66}$ in Eqs.~(\ref{EqA20})-(\ref{EqA26}). Thus, the frequency
of the torsional acoustic mode becomes
\begin{equation}
    {\omega} = \sqrt{\frac{c_{66}}{\rho_{2D}}} k \,,%
\label{EqA27}
\end{equation}
which is identical to Eq.~(\ref{Eq8}). The torsional frequency is
the same as frequency of the shear motion in the equivalent thin
plate~\cite{DT255}.


\subsubsection{\bf\em Bending}

The Euler-Lagrange equation for bending a tube or a plate is given
by~\cite{DT255}
\begin{equation}
    \frac{d}{dt} \left(%
        \frac{{\partial}\mathcal{L}}{{\partial}\frac{du_z}{dt}}
                \right)%
 -  \frac{d^2}{dx^2} \left(%
        \frac{{\partial}\mathcal{L}}{{\partial}\frac{d^2u_z}{dx^2}}
                \right)%
    = 0 \,.
\label{EqA28}
\end{equation}
Inserting the Lagrangian of Eq.~(\ref{EqA19}) for flexural motion
in the Euler-Lagrange Eq.~(\ref{EqA28}) yields the wave equation
for flexual vibrations
\begin{eqnarray}
    {\rho_{1D}}\frac{d^2u_z}{dt^2} %
    + {{\pi}c_{11}R^3} \left( 1+\frac{D}{{c_{11}}R^2} \right)
    \frac{d^4u_z}{dx^4} &=& \nonumber \\
    {\rho_{1D}}\frac{d^2u_z}{dt^2} %
    +D_b \frac{d^4u_z}{dx^4} &=& 0 \,.
\label{EqA29}
\end{eqnarray}
This wave equation can be solved using the ansatz
\begin{equation}
    u_z = u_{z,0} e^{i(kx-{\omega}t)} %
\label{EqA30}
\end{equation}
to yield
\begin{equation}
    {\rho_{1D}}{\omega}^2 = {{\pi}c_{11}R^3}
    \left( 1+\frac{D}{{c_{11}}R^2} \right)
     k^4 = D_b k^4\,.%
\label{EqA31}
\end{equation}
This finally translates to the desired form
\begin{equation}
{\omega}=\sqrt{\frac{{\pi c_{11}R^3}}{\rho_{1D}} %
    \left( 1+\frac{D}{{c_{11}}R^2} \right) }~k^2 %
    = \sqrt{\frac{D_b}{\rho_{1D}}}~k^2\,,%
\label{EqA32}
\end{equation}
which is identical to Eq.~(\ref{Eq9}).

For a liquid-filled nanotube, we only need to replace $\rho_{1D}$
by $\tilde\rho_{1D}$ in Eq.~(\ref{EqA32}) to get
\begin{equation}
{\omega}=\sqrt{\frac{{\pi}c_{11}R^3}{\tilde\rho_{1D}}
    \left( 1+\frac{D}{{c_{11}}R^2} \right) }~k^2 %
    = \sqrt{\frac{D_b}{\tilde\rho_{1D}}}~k^2\,,%
\label{EqA33}
\end{equation}
which is identical to Eq.~(\ref{Eq14}).

\subsection{C. Coupling Between a Travelling Pressure Wave
and the RBM in a Liquid-Filled Carbon Nanotube}

\begin{figure}[t]
\includegraphics[width=1.0\columnwidth]{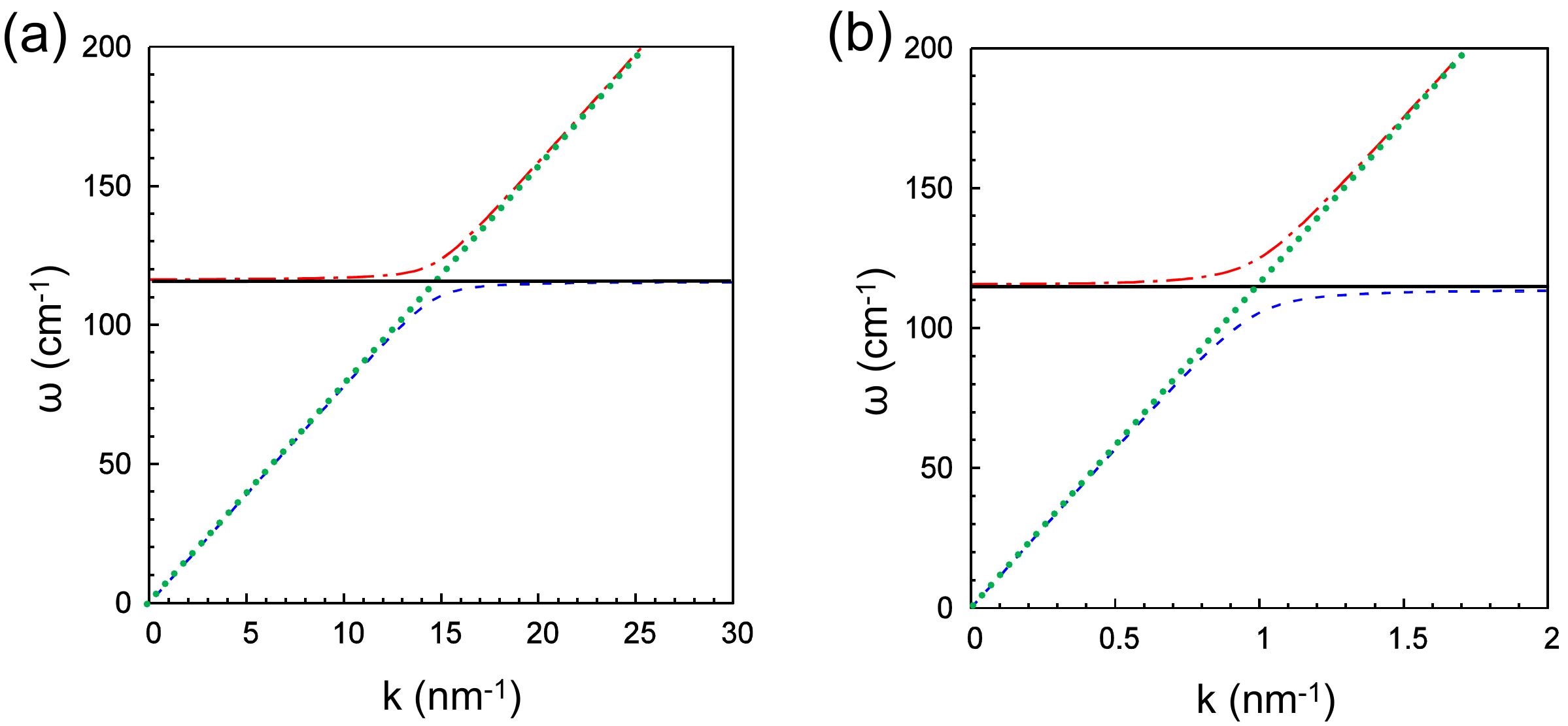}
\caption{(Color online) Nature and coupling of vibration modes
$\omega(k)$ of a carbon nanotube that may be filled with water.
(a) Coupling between the dispersionless RBM of a CNT, shown by the
black line, and the pressure wave of water enclosed in the CNT,
shown by the green dotted line. (b) Coupling between the RBM of a
CNT, shown by the black line, and the longitudinal acoustic mode
of the CNT, shown by the green dotted line. Results are presented
for a CNT with a radius of 1~nm.%
\label{fig4} }
\end{figure}

Next we consider a long-wavelength displacement wave
$u_x=u_{x,0}\exp[i(kx-{\omega}t)]$ of small frequency $\omega$ and
wave vector $k$ travelling down the liquid column filling a carbon
nanotube. We assume the liquid to be compressible but
viscosity-free. Thus, the travelling displacement wave will result
in a pressure wave $p=p_0\exp[i(kx-{\omega}t)]$ that causes radial
displacements $r=r_0\exp[i(kx-{\omega}t)]$ in the CNT wall. These
radial displacements couple the pressure wave in the liquid to the
RBM, but not the longitudinal and torsional modes of the CNT.

At small frequencies $\omega$, there will be little radial
variation in the pressure. The local compressive strain in the
liquid will thus be
\begin{equation}
-\frac{\delta V}{V}=-\left(\frac{\partial u_x}{\partial x}%
+\frac{2 r}{R}\right) %
\label{EqA34}
\end{equation}
and the pressure becomes
\begin{equation}
p=-B \frac{\delta V}{V}=-B \left(\frac{\partial u_x}{\partial x}%
+\frac{2 r}{R}\right) \,. %
\label{EqA35}
\end{equation}
Here, $B$ is the bulk modulus of the filling liquid, which we
assume is water with $B=2.2{\cdot}10^9$~Pa.

The local acceleration of water is given by
\begin{equation}
{\rho}_l \frac{{\partial}^2 u_x}{{\partial}t^2}=-\frac{\partial
p}{\partial x} %
\label{EqA36}
\end{equation}
and the radial acceleration of the CNT is given by
\begin{equation}
{\rho}_{2D} \frac{{\partial}^2 r}{{\partial}t^2} %
= p - \frac{c_{11}}{R^2}r \,. %
\label{EqA37}
\end{equation}
Inserting harmonic solutions for $p$, $u_x$ and $r$ into
Eqs.~(\ref{EqA35})-(\ref{EqA37}), we obtain
\begin{eqnarray}
\mkern-50mu 
\left( \begin{array}{ccc}
 1 & Bik & 2B/R \\
 -ik & {\rho}_l {\omega}^2 & 0 \\
  1     &   0    & -(c_{11}/{R^2}-{\rho}_{2D}{\omega}^2)
       \end{array}
\right)
\mkern-10mu 
\left(
\begin{array}{c}
p_0\\
u_{x, 0}\\
r_0
\end{array}
\right) = 0 %
\label{EqA38}
\end{eqnarray}
with the characteristic equation
\begin{equation}
\left(\frac{c_{11}}{R^2}%
-{\rho}_{2D}{\omega}^2\right)%
\left(k^2B-{\rho}_l{\omega}^2\right)%
-{\rho}_l {\omega}^2\frac{2B}{R} = 0 \,.%
\label{EqA39}
\end{equation}
This can be rewritten as
\begin{equation}
\left({\omega}_0^2-{\omega}^2\right) %
\left(k^2v^2-{\omega}^2\right)-{\omega}^2{\gamma}^2=0 \,,
\label{EqA40}
\end{equation}
where %
${\omega}_0^2={c_{11}}/({\rho}_{2D}R^2)$ and %
${\gamma}^2={2B}/({\rho}_{2D}R)$. %
For a CNT of radius $R=1$~nm, we obtain
${\omega}_0^2=474~$ps$^{-2}$ and %
${\gamma}^2=5.92~$ps$^{-2}$. %

Solving Eq.~(\ref{EqA40}) leads to the dispersion relation
$\omega(k)$, which is presented in Fig.~\ref{fig4}(a). In the
following, we focus on the lowest lying branch of the dispersion
relation describing a long wavelength, low frequency pressure wave
travelling down the liquid column. In this case, we can neglect
$\omega^2$ in the first factor of Eq.~(\ref{EqA40}) and obtain
\begin{equation}
\omega=\frac{vk}{\sqrt{1+\frac{\gamma^2}{\omega_{0}^2}}} \,. %
\label{EqA41}
\end{equation}

Considering the filling liquid to be water with the speed of sound
$v=\sqrt{B/{\rho}_l}{\approx}1483$~m/s, the velocity of the
propagating pressure wave inside the CNT becomes
\begin{equation}
\frac{\omega}{k} = v
\frac{1}{\sqrt{1+\frac{\gamma^2}{\omega_{0}^2}}}
=1474~{\rm m/s} \,.%
\label{EqA42}
\end{equation}
This value is only slightly reduced from that of bulk water
because of the relative rigidity of the CNT.

In the corresponding low wavenumber range, the frequency of the
RBM is changed to
\begin{equation}
\tilde\omega_{RBM}=\omega_{RBM}{\sqrt{1+\frac{\gamma^2}{\omega_{RBM}^2}}}\,.%
\label{EqA43}
\end{equation}
For a CNT with radius $R=1$~nm, $(1+\gamma^2/\omega_{RBM}^2)^{1/2}
= (1+5.92/474)^{1/2}=1.006$, yielding only a $0.6$\% increase in
frequency.

The situation is quite different for tubulin microtubules.
Assuming a radius of $R=12.8$~nm, we find
$\omega_{RBM}=0.24$~cm$^{-1}$ corresponding to
$\omega_{RBM}^{2}=0.00203$~ps$^{-2}$. In that case,
$(1+\gamma^2/\omega_{RBM}^2)^{1/2} =(1+0.0859/0.00203)^{1/2}=6.6$.
In other words, filling tubulin microtubules with water will
increase their RBM frequency by a factor of 6.6.

As seen in the full solution of Eq.~(\ref{EqA40}) in
Fig.~\ref{fig4}(a), at $k{\approx}15$~nm$^{-1}$, there is level
repulsion and interchange in character between the two dispersion
curves. At very much higher frequencies there will be radial modes
in the water column that will couple to the RBM of the CNT. These
lie outside the scope of the present treatment.

\subsection{D. Coupling Between the LA Mode and the RBM
in Carbon Nanotubes}

Consider a longitudinal wave travelling along a CNT containing no
liquid. The CNT of radius $R$ is aligned along the $x-$direction
and can be thought of as a rolled up %
plate in the $xy-$plane with a width of $2{\pi}R$ along the
$y-$direction. Where the CNT is being locally stretched, it will
narrow down and where it is compressed, it will fatten due to the
nonzero value of $c_{12}$ reflected in the Poisson ratio. For
longitudinal displacement $u_x$ and radial displacement $r$, the
strains will be $\epsilon_{11}={\partial u_x}/{\partial x}$ and
$\epsilon_{22}={r}/{R}$.

The strain energy density is then
\begin{equation}
\begin{split}
U &=\frac{1}{2}\left( c_{11}\epsilon_{11}^2%
+c_{11}\epsilon_{22}^2%
+2c_{12}\epsilon_{22}\epsilon_{11}\right) \\
&=\frac{1}{2}c_{11}\left(\left(\frac{\partial u_x}{\partial x}\right)^2%
+\left(\frac{r}{R}\right)^2\right)%
+c_{12}\frac{\partial u_x}{\partial x} \frac{r}{R} %
\label{EqA44}
\end{split}
\end{equation}
and the kinetic energy density is
\begin{equation}
T =\frac{1}{2}{\rho}_{2D}\left(\left(\frac{{\partial}u_x}%
{{\partial}t}\right)^2+%
\left(\frac{\partial r}{\partial t}\right)^2\right) \,. %
\label{EqA45}
\end{equation}
There are two Euler-Langrange equations for the radial and the
axial motion,
\begin{equation}
\frac{d}{dt}\left(\frac{\partial \mathcal{L}}%
{\partial (\frac{dr}{dt})}\right)%
-\frac{\partial \mathcal{L}}{\partial r}=0 %
\label{EqA46}
\end{equation}
and
\begin{equation}
\frac{d}{dt}\left(\frac{\partial %
\mathcal{L}}{\partial (\frac{du_x}{dt})}\right)%
+\frac{\partial}{\partial x}%
\left(\frac{\partial\mathcal{L}}%
{\partial (\frac{du_x}{dx})}\right)=0 \,.%
\label{EqA47}
\end{equation}
With the Lagrangian $\mathcal{L}=T-U$ given by
Eqs.~(\ref{EqA44})-(\ref{EqA45}), the Euler-Lagrange equations
translate to partial differential equations
\begin{equation}
{\rho}_{2D}\frac{\partial^2 r}{\partial t^2}%
+c_{11}\frac{r}{R^2}%
+\frac{c_{12}}{R}\frac{\partial u_x}{\partial x}=0 %
\label{EqA48}
\end{equation}
and
\begin{equation}
{\rho}_{2D}\frac{\partial^2 u_x}{\partial t^2}%
-c_{11}\frac{\partial^2 u_x}{\partial x^2}%
-\frac{c_{12}}{R}\frac{\partial r}{\partial x}=0 \,. %
\label{EqA49}
\end{equation}
Assuming harmonic solutions $u_x=u_{x,0}\exp[i(kx-{\omega}t)]$ and
$r=r_0\exp[i(kx-{\omega}t)]$, we get
\begin{eqnarray}
\left( \begin{array}{cc}
 -{\rho}_{2D}{\omega}^2+c_{11}/R^2 & ikc_{12}/R  \\
- ikc_{12}/R & -{\rho}_{2D}{\omega}^2+c_{11}k^2 )
       \end{array}
\right)
\left(\begin{array}{c}
r_0\\
u_{x, 0}
       \end{array}
\right) =0
\nonumber \\
\label{EqA50}
\end{eqnarray}
with the characteristic equation
\begin{equation}
\left(\frac{c_{11}}{{\rho_{2D}}R^2}-%
{\omega}^2\right)\left(\frac{c_{11}}{{\rho_{2D}}}k^2-%
{\omega}^2\right)-\frac{c_{12}^2k^2}{{\rho_{2D}}^2R^2}=0\,.
\label{EqA51}
\end{equation}

Solving Eq.~(\ref{EqA51}) leads to the dispersion relations
${\omega}(k)$ that are shown in Fig.~\ref{fig4}(b) for a CNT with
radius $R=1$~nm, with the values
${c_{11}}/(\rho_{2D}R^2)=474$~ps$^{-2}$, %
${c_{11}}/(\rho_{2D})=474$~ps$^{-2}$nm$^2$ and
${c_{12}^2}/(\rho_{2D}^2R^2)=6434$~nm$^2$ps$^{-4}$. Our results in
Fig.~\ref{fig4}(b) closely resemble those of
Ref.~\cite{WangMicrotubules}, %
obtained using a more complex formalism describing orthotropic
elastic cylindrical shells using somewhat different input
parameters. Were $c_{12}$ to be zero, then the uncoupled solutions
would be the dispersionless RBM of frequency
\begin{equation}
\omega = \frac{1}{R} \sqrt{\frac{c_{11}}{\rho_{2D}}} \,, %
\label{EqA52}
\end{equation}
shown by the black solid line in Fig.~\ref{fig4}(b), and the pure
longitudinal mode of velocity
\begin{equation}
v = \frac{\omega}{k}=\sqrt{\frac{c_{11}}{\rho_{2D}}} \,, %
\label{EqA53}
\end{equation}
shown by the green dotted line in Fig.~\ref{fig4}(b).

The coupling term induces level repulsion between the
${\omega}_{-}(k)$ and ${\omega}_{+}(k)$ branches, with strong mode
hybridization occurring near $k{\approx}1$~nm$^{-1}$. It is of
interest to examine the limiting forms of the two solutions for
$k{\to}0$ and $k{\to}\infty$.

For $k{\to}0$, the larger solution ${\omega}_+(k)$ approaches a
constant value $\omega_{+}^0$. From Eq.~(\ref{EqA51}) we obtain
\begin{equation}
\omega_{+}^0=\frac{1}{R}\sqrt{\frac{c_{11}}{\rho_{2D}}}+
\mathcal{O}\left(k^2 \right) \,. %
\label{EqA54}
\end{equation}
The lower solution ${\omega}_-(k)$ approaches the value
$\omega_{-}=vk$, where $v$ is the velocity of longitudinal mode,
modified by its coupling to the RBM. Inserting this in
Eq.~(\ref{EqA51}) and taking the limit $k{\to}0$, we obtain
\begin{equation}
v = \frac{\omega}{k}=\sqrt{\frac{c_{11} %
\left( 1-\alpha^2\right)}{\rho_{2D}}} \,,%
\label{EqA55}
\end{equation}
where $\alpha=c_{12}/c_{11}$ is the Poisson ratio. The numerical
value of the velocity obtained using this expression,
$v=21.45$~nm/ps, is slightly smaller than the velocity of the
longitudinal mode
\begin{equation}
v = \frac{\omega}{k}=\sqrt{\frac{c_{11}}{\rho_{2D}}} \,,%
\label{EqA56}
\end{equation}
which turns out to be $v=21.77$~nm/ps. The 1\% reduction by the
factor of $\sqrt{1-\alpha^2}$ is caused by the coupling of the
longitudinal mode to the RBM.

In the opposite limit $k{\to}\infty$, the lower solution
${\omega}_{-}(k)$ approaches a value, which is a little below the
uncoupled RBM frequency~\cite{DT255} of a nanotube with $R=1$~nm,
\begin{equation}
\omega = \frac{1}{R} \sqrt{\frac{c_{11}}{\rho_{2D}}}%
       = 117~{\rm cm}^{-1}  \,. %
\label{EqA57}
\end{equation}
We can obtain the coupled RBM frequency $\omega_{-}^{\infty}$ from
Eq.~(\ref{EqA51}) by neglecting its value in comparison with
$c_{11}k^2/\rho_{2D}$. This yields
\begin{equation}
\omega = \frac{1}{R}\sqrt{\frac{c_{11} %
         \left( 1-\alpha^2\right)}{\rho_{2D}}} %
       = 115~{\rm cm}^{-1} \,. %
\label{EqA58}
\end{equation}
The 1\% reduction of the RMB frequency in the $k{\to}\infty$ limit
by the factor of $\sqrt{1-\alpha^2}$ is again caused by the
coupling of the longitudinal mode to the RBM.

For $k{\to}\infty$, the upper solution approaches the value
$\omega_{+}=vk$, where $v$ is the velocity of the uncoupled LA
mode. Neglecting ${c_{11}}/(\rho_{2D}R)$ in comparison with
$v^2k^2$ and neglecting $k^2$ terms in comparison with $k^4$ terms
in Eq.~(\ref{EqA51}), we arrive at
\begin{equation}
v=\sqrt{{c_{11}}/{\rho_{2D}}} %
\label{EqA59}
\end{equation}
with no corrections due to the coupling to the RBM. This was to be
expected, since for $kR{\gg}1$, this nanotube mode corresponds to
the LA mode in a graphene sheet.


%

\end{document}